\UseRawInputEncoding
\documentclass[aps,prl,reprint,groupedaddress]{revtex4-1}
\usepackage{amsthm,amsmath,amssymb}
\usepackage{graphicx}
\usepackage{epstopdf}
\usepackage{mathrsfs}
\usepackage{float}
\usepackage[colorlinks=true,dvipdfm]{hyperref}

\begin{document}

\title{Effective-Dimension Theory of Critical Phenomena above Upper Critical Dimensions}


\author{Shaolong Zeng, Sue Ping Szeto, Fan Zhong}
\thanks{Corresponding author: stszf@mail.sysu.edu.cn}
\affiliation{State Key Laboratory of Optoelectronic Materials and Technologies, School of Physics, Sun Yat-Sen University, Guangzhou 510275, People's Republic of China}

\date{\today}

\begin{abstract}
Phase transitions and critical phenomena are among the most intriguing phenomena in nature and their renormalization-group theory is one of the greatest achievements of theoretical physics. However, the predictions of the theory above an upper critical dimension $d_c$ seriously disagree with reality. In addition to its fundamental significance, the problem is also of practical importance because both complex systems with spatial or temporal long-range interactions and quantum phase transitions can substantially lower $d_c$. The extant scenarios built on a dangerous irrelevant variable (DIV) to resolve the problem introduce two sets of critical exponents and even two sets of scaling laws whose origin is obscure. Here, we consider the DIV from a different perspective and clearly unveil the origin of the two sets of exponents and hence the intrinsic inconsistency in those scenarios. We then develop an effective-dimension theory in which critical fluctuations and system volume are fixed at an effective dimension by the DIV. This enables us to account for all the extant results consistently. A novel asymptotic finite-size scaling behavior for a correlation function together with a new anomalous dimension and its associated scaling law is also derived.
\end{abstract}


\maketitle
\textit{Introduction.}---Phase transitions and critical phenomena are among the most intriguing phenomena in nature and their renormalization-group (RG) theory is one of the greatest achievements of theoretical physics~\cite{Wilson,Mask,Goldenfeld,Cardyb,Justin,amitb,Vasilev,Stanley}. A remarkable result of the RG theory is that there exists an upper critical dimension $d_c$ that divides a nonclassical regime controlled by a nontrivial fixed point below it from a classical regime governed by Gaussian fixed points above it. However, it is well known that the Gaussian exponents recover, only exactly at $d_c$, the Landau mean-field critical exponents, which are known to be valid for $d\geq d_c$ with possible logarithmic corrections at $d_c$. Moreover, a hyperscaling law
\begin{equation}
2-\alpha=d\nu,\label{Widom}
\end{equation}
predicted by the RG theory to be universally valid is only true precisely at $d_c$, where $\alpha$ and $\nu$ are the usual critical exponents and $d$ denotes the spatial dimensionality. Although the most familiar $d_c$ is $4$ and seems experimental irrelevant, classical spatial~\cite{Fisher} and temporal~\cite{Zeng} long-range interactions can lower $d_c$ even down to $d_c=1$ on the one hand and quantum phase transitions have a $d_c$ lowered by the dynamic critical exponent $z$ due to the quantum-classical correspondence~\cite{Sachdev} on the other hand. Therefore, how to resolve these drawbacks of the RG theory in $d>d_c$ is not only of fundamental significance but also of practical relevance and thus has attracted continuously attention~\cite{Brezin82,Brezin85,Privman,Chen,Chen0,Cara,Binder,BNPY,Binder87,Rickwardt,Mon,Parisi,Luijten96,Blote,Luijten97,Luijten99,Aktekin,Aktekin00,Aktekin1,Binder01,Merdan,Merdan5,Jones,Merdan06,Kenna14,Kenna15,Flores15,Flores,Grimm,Zhou,Fang,Lv,Rudnick,Lundow,Lundow14,Wittman,Lundow16,Lundow21,Berche,Kenna13}.

Several main scenarios have been proposed. An early scenario is to consider a dangerous irrelevant variable (DIV)~\cite{Fisherb} that is believed to be responsible for the violation of the hyperscaling law~\cite{Binder,BNPY,Binder87}. This leads to new RG eigenvalues that both bring the Gaussian exponents to their mean-field values and yield special finite-size scaling (FSS)~\cite{fss,Barber,Cardy,Privmanb,Brankov} exponents in agreement with analytical and numerical results~\cite{Brezin82,Brezin85,Privman,Chen,Chen0,Cara,Binder,BNPY,Binder87,Rickwardt,Mon,Parisi,Luijten96,Blote,Luijten99,Aktekin,Aktekin00,Aktekin1,Binder01,Merdan,Merdan5,Jones,Merdan06,Kenna14,Kenna15,Flores15,Rudnick,Wittman,Berche,Kenna13}. However, a correlation length has to be bounded by a system length and new length scales have to be introduced~\cite{Binder,Binder87}. Another scenario is based on the numerical result that the correlation length scales with the system length with an exponent $q=d/d_c$ for $d>d_c$ at criticality~\cite{Jones} (which breaks the bound) and introduces $q$, and hence Q-scaling, as a new exponent. This enables one to study consistently the effects of the DIV~\cite{Kenna14,Kenna15,Flores15,Flores,Berche,Kenna13}. Yet another recent scenario is to combine both the Gaussian and the mean-field exponents linearly~\cite{Lv} from geometric reasons~\cite{Grimm,Zhou,Fang}.

Here, we first show that there still exist several pitfalls in the Q-scaling. Moreover, the two scenarios based on the DIV introduce two sets of critical exponents and even two sets of scaling laws whose origin is obscure. We then consider the DIV from a different perspective and clearly unveil the origin of the two sets of exponents and thus the intrinsic inconsistency in the two scenarios. Finally, we develop an effective-dimension theory in which critical fluctuations and system volume are fixed at an effective dimension by the DIV. This enables us to account for all the extant results consistently as well as to predict new ones.

\textit{Hamiltonian and its exponents.}---Consider a usual $\phi^4$ theory of the critical phenomena described by
\begin{equation}
 \mathcal{H}=\!\int\!{d^dx}\left\{\frac{1}{2}\tau\phi^2+\frac{1}{2}
		\left(\nabla_x^{\frac{\sigma}{2}}\phi\right)^2-h\phi+\frac{1}{4!}u\phi^4\right\},\label{hscrss}
\end{equation}
for an order-parameter field $\phi$, where $\tau$, $u$, and $h$ denote a reduced temperature, a coupling constant, and an ordering field, respectively. For simplicity, we have employed a fractional power $\sigma/2$ for the gradient to represent a spatial long-range interaction algebraically decaying with an exponent $d+\sigma$~\cite{Samko,Metzler,Zaslavsky,West,Tarasovb,Tarasov}. In this way, short-range interaction is covered directly by $\sigma=2$. Given that $\mathcal{H}$ is dimensionless and $[x]=-1$~\cite{Justin,amitb}, the engineering dimensions of other quantities, denoted by square brackets, are
\begin{equation}\label{dimsi}
[\tau]=\sigma,~[\phi]=\frac{d-\sigma}{2},~[h]=\frac{d+\sigma}{2},~[u]=2\sigma-d,
\end{equation}
from which $d_c=2\sigma$ where $[u]=0$. Since the correlation length $\xi\sim|\tau|^{-\nu}$ asymptotically, one has $|\tau|\sim\xi^{-1/\nu}$. Accordingly, the order parameter $M\sim|\tau|^{\beta}\sim\xi^{-\beta/\nu}$ and $h\sim M^{\delta}\sim\xi^{-\beta\delta/\nu}$, where $\beta$ and $\delta$ are critical exponents. These indicate that the critical exponents are given by,
\begin{equation}
\nu=1/[\tau], \quad\beta/\nu=[\phi],\quad \beta\delta/\nu=[h].\label{expdim}
\end{equation}
As a result, the Gaussian exponents are given by
\begin{equation}
\begin{split}
\nu=1/\sigma,\quad\beta=(d-\sigma)2\sigma,\quad\delta=(d+\sigma)/(d-\sigma),\\
\alpha=(2\sigma-d)/\sigma,\quad\gamma=1,\quad\eta=2-\sigma,\qquad\;\label{gauss}
\end{split}
\end{equation}
using Eq.~\eqref{dimsi} and the scaling laws~\cite{Mask,Goldenfeld,Cardyb,Justin,amitb,Vasilev,Stanley},
\begin{equation}\label{scalinglaw}
\alpha+2\beta+\gamma=2,\quad\beta(\delta-1)=\gamma,\quad(2-\eta)\nu=\gamma,
\end{equation}
while the Landau mean-field exponents are,
\begin{equation}
\nu=1/\sigma,~\beta=1/2,~\delta=3,~\alpha=0,~\gamma=1,~\eta=2-\sigma,\label{landauo}
\end{equation}
which are the Gaussian exponents~\eqref{gauss} at $d=d_c$ and thus violate the hyperscaling law~\eqref{Widom}, where $\eta$ and $\gamma$ are also standard critical exponents.

\textit{DIV and Q-scaling.}---We now briefly review the two scenarios uniformly in terms of the engineering dimensions instead of the RG eigenvalues~\cite{BNPY,Luijten96,Luijten97,Flores15}, although they are equal for $d\geq d_c$. These scenarios start with the scaling hypothesis for the singular part of the free energy~\cite{Mask,Goldenfeld,Cardyb,Justin,amitb,Vasilev,Stanley,Fisherb},
\begin{equation}
f(\tau,h,u)=b^{-d}f(\tau b^{[\tau]}, hb^{[h]},ub^{[u]}),\label{fthu}
\end{equation}
where $b$ is a length rescaling factor. $u$ is irrelevant due to $[u]<0$ for $d>d_c$ but dangerous in the sense that $f$ is singular for $u\rightarrow0$. Assuming multiplicative singularity, one writes~\cite{BNPY,Luijten96,Luijten97,Flores15}
\begin{eqnarray}
f(\tau,h,u)&=&b^{-d}f(\tau b^{[\tau]}(ub^{[u]})^{-1/2}, hb^{[h]}(ub^{[u]})^{-1/4})\nonumber\\
&\equiv&b^{-d}f(\tau b^{[\tau^*]}/u^{1/2}, hb^{[h^*]}/u^{1/4}),\label{fthus}
\end{eqnarray}
where
\begin{equation}
[\tau^*]\equiv[\tau]-\frac{1}{2}[u]=\frac{d}{2},\quad [h^*]\equiv[h]-\frac{1}{4}[u]=\frac{3d}{4},\label{taust}
\end{equation}
using Eq.~\eqref{dimsi}. Moreover, $M=-(\partial f/\partial h)_{\tau}$ yields
\begin{equation}
[\phi^*]\equiv[\phi]+[u]/4=d-[h^*]=d/4,\label{mst}
\end{equation}
with Eqs.~\eqref{fthus} and~\eqref{dimsi}. Accordingly, one obtains from Eq.~\eqref{expdim},
\begin{equation}
\nu^*=2/d,\quad \beta=1/2,\quad \delta=3,\label{landau}
\end{equation}
the last two exponents of which return correctly to the Landau forms~\eqref{landauo}, though one introduces an additional $\nu^*$.

In FSS, one inserts one more scaled variable $L^{-1}b$ into the arguments of $f$ on the right-hand side of Eq.~\eqref{fthus} and then let $b=L$, the system length, leading to
\begin{equation}
f(\tau,h,u,L)=L^{-d}f(\tau L^{[\tau^*]}/u^{1/2}, hL^{[h^*]}/u^{1/4}).\label{fthul}
\end{equation}
Accordingly, one finds for $M$ and the susceptibility $\chi=(\partial M/\partial h)_{\tau}$,
\begin{equation}
M\sim L^{-d/4},\quad \chi\sim L^{d/2},\label{mxl}
\end{equation}
at the criticality of $\tau=0$ and $h=0$, in agreement with numerical results~\cite{Binder,BNPY,Rickwardt,Mon,Parisi,Luijten96,Luijten97,Blote,Luijten99,Aktekin,Aktekin00,Aktekin1,Binder01,Merdan,Merdan5,Flores15}, rather than the usual Landau mean-field results $M\sim L^{-\beta/\nu}=L^{-\sigma/2}$ and $\chi\sim L^{\gamma/\nu}=L^{\sigma}$, viz., Eq.~\eqref{mxl} at $d_c=2\sigma$ precisely.

The Q-scaling is a generalization of the DIV to the correlation length and correlation function~\cite{Kenna14,Kenna15,Flores15,Flores,Berche,Kenna13}. One writes the FSS of $\xi$ as
\begin{equation}
\xi(\tau,h,L)=L^q\xi(\tau L^{[\tau^*]}/u^{1/2}, hL^{[h^*]}/u^{1/4}),\label{xilq}
\end{equation}
in agreement with numerical results~\cite{Jones}, with
\begin{equation}\label{qdc}
q=\left\{
\begin{array}{ll}
d/d_c, & {\rm for}~d\geq d_c,\\
1, & {\rm for}~d<d_c,
\end{array}
\right.
\end{equation}
so that $\xi\sim L^q>L$ for $d>d_c$. This results in $\xi^{d_c}\sim L^d$,
which indicates that the volume associated with $\xi$ in $d_c$-dimensions corresponds to actual volume of the system~\cite{Berche}. From Eq.~\eqref{xilq}, in the thermodynamic limit in which $L\rightarrow\infty$, one sets $L=\tau^{-1/[\tau^*]}$ and $L=h^{-1/[h^*]}$ and finds $\xi\sim \tau^{-q/[\tau^*]}$ and $\xi\sim h^{-q/[h^*]}$, respectively. Similar consideration leads to $\xi\sim M^{-q/[\phi^*]}$. Comparing these with the case of $q=1$, one must have
\begin{eqnarray}
q[\tau]=[\tau^*],\qquad\quad\label{qtst}\\
q[h]=[h^*], \quad q[\phi]=[\phi^*],\label{qhst}
\end{eqnarray}
for consistency. Note however that although Eq.~\eqref{qtst} is correct for all $d>d_c=2\sigma$ from Eqs.~\eqref{dimsi} and~\eqref{taust}, the two equalities in Eq.~\eqref{qhst} are only valid for $d=d_c$~\cite{Kenna17} from Eqs~\eqref{dimsi},~\eqref{taust} and~\eqref{mst}.

As one has two different $\nu$, according to the scaling law~\eqref{scalinglaw}, one can have
\begin{eqnarray}
\eta=2-\gamma/\nu=2-\sigma,\quad\label{etao}\\
\eta^*=2-\gamma/\nu^*=2-d/2,\label{etast}
\end{eqnarray}
since $\gamma=1$ from Eqs.~\eqref{landau} and~\eqref{scalinglaw}. This implies that one has, in additional to the standard scaling law~\eqref{scalinglaw},
\begin{eqnarray}
(2-\eta^*)=\gamma/\nu^*=q\gamma/\nu,\:\:\label{fisherlst}\\
2-\alpha=d\nu^*=d\nu/q=d_c\nu,\label{Widomq}
\end{eqnarray}
where the second is the modified hyperscaling law which is now valid for $d\ge d_c$. A negative $\eta^*$ for $d>d_c$ has been estimated~\cite{Nagle} and analytical results show that $\eta$ controls short long-range order while $\eta^*$ long long-range order~\cite{Baker}. Only exactly at $d_c=2\sigma$ do both sets of the exponents coincide.

Equation~\eqref{xilq} introduces to the system two length scales, one is the system length $L$ and the other $\xi$~\cite{Kenna14,Kenna15}. This was employed to account for the two $\eta$ forms and particularly the negative $\eta^*$ for $d>d_c$, which appears to be forbidden by first principles~\cite{Fisher64,Fisher69,Delam}. If distance is measured on the scale of $L$, since the correlation function $G(r,\tau,u)$ scales as
\begin{eqnarray}
G(r,\tau,u)&=&b^{-2[\phi^*]}G(rb^{-1},\tau b^{[\tau^*]}/u^{1/2})/u^{1/2},\label{gcf}\\
&\sim&r^{-2[\phi^*]}=r^{-d/2}, \label{grd2}
\end{eqnarray}
where we have set $b=r$, the distance between two points, in Eq.~\eqref{grd2}. Together with the definition of $\eta$ by
\begin{equation}
G(r,0,u)\sim r^{-d+2-\eta},\label{grd2e}
\end{equation}
Eq.~\eqref{grd2} leads to the negative $\eta^*$, Eq.~\eqref{etast}. However, if distance is measured on the scale of $\xi$, one assumes that Eq.~\eqref{gcf} is replaced by~\cite{Kenna14,Kenna15}
\begin{eqnarray}
G(r,\tau,u)&=&b^{-2[\phi^*]}G(rb^{-q},\tau b^{[\tau^*]}/u^{1/2})/u^{1/2},\label{gcfq}\\
&\sim&  r^{-2[\phi^*]/q}=r^{-d_c/2}=r^{-\sigma},\label{grdc2}
\end{eqnarray}
and hence
\begin{equation}
\eta=2+d_c/2-d=2+\sigma-d,\label{e2ddc2}
\end{equation}
which becomes $\eta=2-\sigma$, Eq.~\eqref{etao}, only exactly at $d_c=2\sigma$~\cite{Kenna15}. Two different decaying regimes of the correlation function have been observed numerically~\cite{Luijten97}. Therefore, the Q-scaling appears to correctly account for the critical behavior above $d_c$, though Eq.~\eqref{qhst} is only true at $d=d_c$ and the same condition is also  mandatory for Eq.~\eqref{e2ddc2} to recover the standard law~\eqref{etao}. Moreover, it is quite strange how can the same $r$ appear sometimes as $rb^{-1}$ and sometimes as $rb^{-q}$ in the same function $G$.

\textit{Origin of the two sets of the exponents.}---To unveil the origin of the two sets of the exponents and scaling laws, we now look at the DIV in a different perspective. We make a transformation~\cite{Brezin85,Luijten96,Luijten97}
\begin{equation}
\phi^*=\phi u^{\frac{1}{4}},~h^*=h u^{-\frac{1}{4}},~\tau^*=\tau u^{-\frac{1}{2}},~x^*=xu^{\frac{1}{2\sigma}},\label{pu4}
\end{equation}
to Eq.~\eqref{hscrss}. This changes it to
\begin{equation}
 \mathcal{H}=\!\int\!{d^dx}\left\{\frac{1}{2}\tau^*\phi^{*2}+\frac{1}{2}
		\left(\nabla_{x^*}^{\frac{\sigma}{2}}\phi^*\right)^2-h^*\phi^*+\frac{1}{4!}\phi^{*4}\right\},\label{hscrs4}
\end{equation}
which contains no $u$ and hence no DIV at all. According to Eq.~\eqref{pu4}, the dimensions of all pertinent quantities changes from Eq.~\eqref{dimsi} to
\begin{eqnarray}\label{dimsu}	
[x^*]&=&[x]+\frac{1}{2\sigma}[u]=-\frac{d}{d_c},\quad[\tau^*]=[\tau]-\frac{1}{2}[u]=\frac{d}{2},\nonumber\\
\protect[\phi^*]&=&[\phi]+\frac{1}{4}[u]=\frac{d}{4},\quad[h^*]=[h]-\frac{1}{4}[u]=\frac{3d}{4}.
\end{eqnarray}
Equation~\eqref{dimsu} implies $\tau^*\sim x^{-[\tau^*]}$ or $x\sim\tau^{*-1/[\tau^*]}$. So,
\begin{equation}
\xi\sim x^{-[x^*]}\sim\tau^{*[x^*]/[\tau^*]}\sim \tau^{*-\nu}.\label{xixp}
\end{equation}
Consequently, Eq.~\eqref{expdim} now becomes
\begin{equation}
\nu=-\frac{[x^*]}{[\tau^*]}, \quad\beta/\nu=-\frac{[\phi^*]}{[x^*]},\quad \beta\delta/\nu=-\frac{[h^*]}{[x^*]},\label{expdimu}
\end{equation}
which correctly returns to Eq.~\eqref{expdim} for $[x^*]=-1$. The Landau mean-field exponents~\eqref{landauo} without the new $\nu^*$ ensue from Eqs.~\eqref{expdimu} and~\eqref{dimsu}.

One sees from Eq.~\eqref{hscrs4} that the transformation~\eqref{pu4} changes all quantities to their starred counterparts~\cite{noteint}, whose dimensions coincide with those in DIV, Eqs.~\eqref{taust} and~\eqref{mst}, except for $x^*$. This strongly hints at the origin of the latter. The Landau exponents describe the critical behavior in the starred quantities, which we refer to as starred space as opposed to the original space, whose critical exponents are the Gaussian ones. Note, however, that the starred exponents do not belong to either of the spaces. How, then, do they come from? In fact, this is simple. They come out from Eq.~\eqref{dimsu} if we set $[x^*]=1$ in Eq.~\eqref{expdimu}. This condition implies that we measure the quantities in the starred space utilizing the length scale $x$ of the original space rather than consistently the starred length scale $x^*$, the central difference between the two spaces. Indeed, from Eq.~\eqref{xixp}, $\xi\sim x^{-[x^*]}$, indicates that in the length scale of the original space, say, $L$, $\xi\sim L^{-[x^*]}\sim L^q$, which underlies the Q-scaling. Therefore, the expressions such as Eqs.~\eqref{fthus} and~\eqref{gcf} in which the starred exponents are employed have based inconsistently on length scales of the different spaces. So have the two different appearances of $r$, $rb^{-1}$ and $rb^{-q}$ in Eqs.~\eqref{gcf} and~\eqref{gcfq}.

Although this perspective produces the Landau mean-field exponents, these exponents cannot account for the peculiar FSS and the violation of the hyperscaling law if we stay consistently in the starred space. Therefore, we need a new theory.

\textit{Effective-dimension theory.}---Instead of Eq.~\eqref{pu4}, we now make another transformation~\cite{Privman},
\begin{equation}
\phi'=\phi u^{1/2},\quad h'=h u^{1/2}\label{pu2}
\end{equation}
to Eq.~\eqref{hscrss}. This leads to
\begin{equation}
 \mathcal{H}=\!\int\!{\left(d^dx u^{-1}\right)}\left\{\frac{1}{2}\tau\phi'^{2}+\frac{1}{2}
		\left(\nabla_{x}^{\frac{\sigma}{2}}\phi'\right)^2-h'\phi'+\frac{1}{4!}\phi'^{4}\right\},\label{hscrs2}
\end{equation}
which contains only an overall $u^{-1}$ factor, which, in turn, diverges at $u\rightarrow0$ and reflects its danger~\cite{Privman}. As a result of Eq.~\eqref{pu2}, we now have
\begin{equation}\label{dimsu2}	
[\phi']=[\phi]+\frac{1}{2}[u]=\frac{d_c}{4},\quad[h']=[h]+\frac{1}{2}[u]=\frac{3d_c}{4},
\end{equation}
using Eq.~\eqref{dimsi}. The dimensions of the length $x$ and $\tau$ remain intact. So does $\nu$. Therefore, we again recover the Landau exponents Eq.~\eqref{landauo} using Eq.~\eqref{expdim}.

The central idea of the theory is that the overall $u^{-1}$ factor and $d^dx$ in Eq.~\eqref{hscrs2} conspire to change the effective dimension of the system to~\cite{Zeng}
\begin{equation}
d_{\rm eff}=d+[u]=d_c,\label{deffsu}
\end{equation}
using Eq.~\eqref{dimsi} since $[d^dx u^{-1}]=-d_{\rm eff}$ similar to the transformed dimensions in Eqs.~\eqref{dimsu} and~\eqref{dimsu2}. This implies that critical fluctuations of a system in a $d>d_c$ dimensional space are fixed at $d_c$, somehow resembling $\xi^{d_c}\sim L^d$ in the Q-scaling and indicates that the DIV now serves also to correct the spatial dimension besides the dimensions, Eq.~\eqref{dimsu2}. Accordingly, the hyperscaling law, Eq.~\eqref{Widom}, must be observed in $d_{\rm eff}$~\cite{Zeng} and becomes Eq.~\eqref{Widomq} and thus holds even for $d>d_c$. Note, however, that $u^{-1}$ does not change $x$. Rather, it can be entirely absorbed in a transformed volume $L'^d=L^d u^{-1}$, since the integration becomes a volume at the tree level for the zero wave-number mode that is responsible for the finite-size behavior~\cite{Brezin85}. Consequently,
\begin{equation}
[L']=-1-[u]/d=-d_c/d=-1/q,\label{ldcd}
\end{equation}
different from $[L]$!

Therefore, we sit now in a core space in which fluctuations and the volume are fixed consistently at $d_{\rm eff}$ of the original $d$-dimensional space by the DIV with original $x$ and $\tau$ and so,
\begin{eqnarray}
f(\tau,h',L')&=&b^{-d_c}f(\tau b^{[\tau]}, h'b^{[h']},L'^{-1}b^{-[L']}),\label{fthsb}\\
\xi(\tau,h',L')&=&b\xi(\tau b^{[\tau]}, h'b^{[h']},L'^{-1}b^{-[L']}),\label{xile}\\
G(r,\tau,L')&=&b^{-2[\phi']}G(rb^{-1},\tau b^{[\tau]}, L'^{-1}b^{-[L']}).\label{grtle}
\end{eqnarray}
Now the two different length scales of the Q-scaling appear naturally but in different places. Their respective FSS forms are
\begin{eqnarray}
f(\tau,h',L')&=&L'^{-d}f(\tau L'^{q[\tau]}, h'L'^{q[h']},1),\label{fthsl}\\
\xi(\tau,h',L')&=&L'^q\xi(\tau L'^{q[\tau]}, h'L'^{q[h']},1),\label{xilel}\\
G(r,\tau,L')&=&L'^{-2q[\phi']}G(rL'^{-q},\tau L'^{q[\tau]},1),\label{grtlel}
\end{eqnarray}
the first two of which are just the Q-scaling FSS forms Eqs.~\eqref{fthul} and~\eqref{xilq} in the absence of $u$ because of Eq.~\eqref{qtst} and
\begin{equation}
q[h']=[h^*],\quad q[\phi']=[\phi^*],\label{qthe}
\end{equation}
using Eqs.~\eqref{dimsu2},~\eqref{taust}, and~\eqref{ldcd}. Equation~\eqref{qthe} is now exact in all $d$ in contrast to Eq.~\eqref{qhst} which is only true at $d_c$. As a result of Eq.~\eqref{fthsl}, Eq.~\eqref{mxl} follows, which agrees with numerical results~\cite{Binder,BNPY,Rickwardt,Mon,Parisi,Luijten96,Luijten97,Blote,Luijten99,Aktekin,Aktekin00,Aktekin1,Binder01,Merdan,Merdan5,Flores15}. Also, from Eqs.~\eqref{grtlel} and~\eqref{dimsu2}, one finds, instead of Eq.~\eqref{grd2},
\begin{equation}
G(r,0,L')=L'^{-2q[\phi']}G(r/L'^q,0,1)\sim L'^{-d/2},\label{gld2}
\end{equation}
at criticality, which is directly supported by numerical results~\cite{Luijten97,Flores15}. Moreover, Eq.~\eqref{grtle} together with Eqs.~\eqref{dimsu2} and~\eqref{ldcd} yields
\begin{equation}
G=r^{-d_c/2}G(1,0,r^{1/q}/L')=r^{-d_c/2}\tilde{G}(r/L'^q),\label{grs0}
\end{equation}
which is somehow similar to Eq.~\eqref{grdc2} except for the scaling function $\tilde{G}$.

We see from Eqs.~\eqref{gld2} and~\eqref{grs0} that we have reproduced the special behaviors of the correlation function in the Q-scaling. The result~\eqref{gld2} is more reasonable than Eq.~\eqref{grd2} since large-scale finite-size behavior is concerned. Indeed, in verification of the leading FSS behavior of Eq.~\eqref{grd2}, one simply set $r=L/2$~\cite{Luijten97,Flores15}. One might also expect the same conclusions for $\eta$ from Eqs.~\eqref{gld2} and~\eqref{grs0}. However, as the scaling form, Eq.~\eqref{grtle} is different, different results emerge.

To see this, we consider the scaling law for $\eta$. According to Eq.~\eqref{grtle}, we ought to change Eq.~\eqref{grd2e} to
\begin{equation}
G(r,0,L')=r^{-d+2-\eta}g(r/L'^q),\label{grd2elq}
\end{equation}
to account for the different scales of $r$ and $L'$. As fluctuations and volume are fixed at $d_{\rm eff}$, we should write~\cite{Kenna14}
\begin{equation}\label{xgl}
\chi\sim\!\int_0^{L'^q}\!{d^{d_{\rm eff}}x}G(r,0,L')\sim L'^{-q(d_c-d+2-\eta)},
\end{equation}
instead of the usual $d$-dimensional integration. Using Eq.~\eqref{mxl}, we arrive at a modified scaling law
\begin{equation}
(d_c-d+2-\eta)\nu=\gamma,\label{fisherd1}
\end{equation}
instead of the standard one~\eqref{scalinglaw}. Accordingly, we find $\eta$ given by Eq.~\eqref{e2ddc2}. Only if $d_c$ were replaced by $d$ in Eq.~\eqref{xgl} could one reach the standard Landau mean-field result~\eqref{etao}. Therefore, Eq.~\eqref{e2ddc2} uniquely arises from the effective dimension and is, instead of Eq.~\eqref{etao}, the appropriate $\eta$ for the behavior~\eqref{grs0}, though they equal at $d_c$.

Moreover, the standard FSS of $G\sim L^{-d+2-\eta}$ must change to
\begin{equation}
G\sim L'^{-q(d-2+\eta)},\label{grd2eq}
\end{equation}
consistently for $r=L'^q$ in Eq.~\eqref{grd2elq}. Comparing Eq.~\eqref{grd2eq} with Eq.~\eqref{gld2}, we obtain again Eq.~\eqref{e2ddc2} in conformity with that resulted from Eq.~\eqref{grs0}.

Therefore, we have shown that the critical behavior above $d_c$ is described uniquely by one special $\eta$, Eq.~\eqref{e2ddc2}, rather than a standard one, Eq.~\eqref{etao}, and the peculiar one, Eq.~\eqref{etast}. Its related scaling law is also modified to Eq.~\eqref{fisherd1} from Eqs.~\eqref{etao} and~\eqref{etast} or~\eqref{fisherlst}.

Two remarks are necessary here. Firstly, the leading behavior, $r^{-d_c/2}=r^{-\sigma}$, in Eq.~\eqref{grs0}, coinciding with the Q-scaling result~\eqref{grdc2}, is different from the usual Gaussian $r^{-d+\sigma}$ analytically derived in an Ising model with exact recursion relations and attributed to be controlled by short long-range correlations~\cite{Baker}. Indeed, Gaussian behavior for $\sigma=2$ is also observed in certain Fourier modes~\cite{Flores}. Secondly, the argument of $\tilde{G}$ in Eq.~\eqref{grs0} ought to be $r/L'^q$ instead of $r/L'$ as was employed in the numerical investigation~\cite{Luijten97}. In fact, the same argument appears in Eq.~\eqref{gld2}. So, there ought to be a small correction of $L'^{1-q}=L'^{-(d-d_c)/d_c}$ for the numerical results at $r=L'/2$~\cite{Luijten97,Flores15}.

Concluding, we have proposed the effective-dimension theory that does produce all the Landau mean-field exponents and remedy the hyperscaling law for $d\geq d_c$. Its central physics is that critical fluctuations and system volume are fixed at the effective dimension by the DIV. The theory circumvents the intrinsic inconsistency of employing different spaces to compute exponents in the previous scenarios, removes the specific requirement for some equalities to hold, and leads to a novel asymptotic FSS behavior for the correlation functions together with a new anomalous dimension and its associated scaling law, as well as results in all extant FSS results.

\begin{acknowledgments}
This work was supported by the National Natural Science Foundation of China (Grant Nos. 11575297 and 12175316).
\end{acknowledgments}


\begin{thebibliography}{99}
\bibitem{noteint} The integration can also be written as $u^{-d/d_c}\!\int{d^dx^*}$ in terms of the starred coordinates. According to the effective-dimension theory below, this amounts to an effective dimension of $d_{\rm eff}=-d[u]/d_c+d[x^*]=-d$, which is just the dimension of $d^dx$ as it ought to be. Therefore, the integration has no real effect.
\end{thebibliography}
\end{document}